\documentclass[aps,twocolumn,pra,tightenlines,floatfix,showpacs,superscriptaddress]{revtex4-1}
\usepackage{graphicx}
\usepackage{amsmath}
\usepackage{amssymb}
\usepackage{times}
\usepackage[english]{babel}
\usepackage{natbib}
\usepackage{hyperref}

\newcommand{\ket}[1]{\ensuremath{\left|#1\right\rangle}}
\newcommand{\bracket}[2]{\ensuremath{\left\langle #1 \middle| #2 \right\rangle}}
\newcommand{\matrixel}[3]{\ensuremath{\left\langle #1 \middle| #2 \middle| #3 \right\rangle}}

\begin{document}
\title{Hysteresis of noninteracting and spin-orbit coupled atomic Fermi gases with relaxation}
\author{Mekena Metcalf}
\affiliation{School of Natural Sciences, University of California, Merced, CA 95343, USA}
\author{Chen-Yen Lai}
\affiliation{School of Natural Sciences, University of California, Merced, CA 95343, USA}
\author{Chih-Chun Chien}
\affiliation{School of Natural Sciences, University of California, Merced, CA 95343, USA}
\email{cchien5@ucmerced.edu}

\begin{abstract}
Hysteresis can be found in driven many-body systems such as magnets and superfluids. Rate-dependent hysteresis arises when a system is driven periodically while relaxing towards equilibrium. A two-state paramagnet driven by an oscillating magnetic field in the relaxation approximation clearly demonstrates rate-dependent hysteresis. A noninteracting atomic Fermi gas in an optical ring potential, when driven by a periodic artificial gauge field and subjected to dissipation, is shown to exhibit hysteresis loops of atomic current due to a competition of the driving time and the relaxation time. This is in contrast to electronic systems exhibiting equilibrium persistent current driven by magnetic flux due to rapid relaxation. Universal behavior of the dissipated energy in one hysteresis loop is observed in both the magnetic and atomic systems, showing linear and inverse-linear dependence on the relaxation time in the strong and weak dissipation regimes. While interactions in general invalidate the framework for rate-dependent hysteresis, an atomic Fermi gas with artificial spin-orbit coupling exhibits hysteresis loops of atomic currents. Cold-atoms in ring-shape potentials are thus promising in demonstrating rate-dependent hysteresis and its associated phenomena.
\end{abstract}


\maketitle

\section{Introduction}

Experimental advancements of cold atoms in optical potentials have renewed interest in the study of persistent current in mesoscopic metallic rings. A one-dimensional (1D) ring of electrons in the presence of magnetic flux can support a non-vanishing current in thermal equilibrium, known as persistent current, and is identical to magnetization \cite{PC_Copper, MesoPhys_Akkermans,NazarovBook}. Recent work has focused on persistent current measurement of atomic superfluids. While bosonic atoms are commonly implemented, studies of fermion persistent current in cold atoms remain less explored \cite{BEC_WeakLink, SF_Hysteresis, Minguzzi_BosonPC}. Past experiments studying persistent current in mesoscopic rings came across difficulties with disorder and interaction effects, and cold atoms seem to be ideal systems for studying the persistent current of non-interacting fermions in a ring~\cite{PC_Copper, PC_GoldLoop}. 

However, a lack of relaxation mechanisms for noninteracting atomic Fermi gases will be detrimental to the experimental measurement of a genuine persistent current because the very definition of a persistent current requires the system to reach thermal equilibrium when magnetic flux is introduced \cite{MesoPhys_Akkermans,NazarovBook}. Without any dissipation mechanism to bring a clean noninteracting atomic Fermi gas back to equilibrium in the presence of an effective magnetic flux, the current induced by an artificial vector potential in a ring cannot be identified as the persistent current. For cold-atom systems, dissipation must be added for them to settle into equilibrium. Exchange of energy with a thermal bath relaxes the system and allows for the persistent current at a fixed value of the magnetic flux to be observed in the long-time limit.

Interestingly, dissipation is not only necessary for the observation of persistent current, but also induces  hysteresis when the artificial magnetic flux is periodically modulated. Hysteresis has been intensely studied in magnetic systems. Well known forms of hysteresis include ferromagnetic materials \cite{MagHyst} and superfluids \cite{SF_Hysteresis}, and it is known to be a significant mechanism in the advancement of modern electronics because it provides memory to a system and can cancel systematic effects~\cite{SFhyst_Theory, Mag_Alloys, MDV_CircMem}. Hysteresis arising from thermal relaxation occurs in two forms: rate-dependent hysteresis and rate-independent hysteresis \cite{MagHyst}. In rate-independent hysteresis, two or more metastable energy states are separated by an energy barrier and the free energy takes a non-linear form. When an external driving force moves the system from one metastable state to another, the system exhibits history-dependent behavior. Rate-independent hysteresis of supercurrent in a rotating, superfluid Bose-Einstein Condensate (BEC) was observed and has been considered a milestone in the advancement of atomtronic circuitry \cite{SF_Hysteresis}. Local energy minima are dependent on the quantized angular momentum of the superfluid and the system can move around different minima by stirring the atoms with an optical potential. After reaching a critical rotation, the angular winding number changes, leading to hysteresis loops of angular velocity. In addtion, a recent experiment has demonstrated hysteresis when a BEC is placed in a double-well potential with tunable interactions and potential minima \cite{DoubleWellHyst}.

In contrast, rate-dependent hysteresis does not require symmetry-broken phases or even energy minima. Instead, the hysteresis arises due to a competition between the time scales of driving and response. Here we explore rate-dependent hysteresis in noninteracting atomic fermions driven by an artificial gauge field. One significant point of using cold atoms is to provide a means to explore the rate-dependent hysteresis outside of magnetic systems. The competition between the period of driving and relaxation time leads to hysteresis loops of mass current when a periodically modulated artificial, magnetic flux and dissipation are present.  Moreover, we will discuss techniques for realizing the setup. 

Universal behavior of dissipated energy in one hysteresis loop, which can be measured by the hysteresis-loop area, will be presented. The area of a hysteresis loop determines how much energy is transferred from the driving force to the reservoir via the system. Rate-independent hysteresis of ferromagnetic materials provides a common illustration of energy dissipation from  hysteresis behavior, and knowledge of the energy dissipation is highly important when ferromagnetic and ferroelectric materials are used in electronic systems \cite{FerroHystTheory}. The ability to form and magnetize domains depends on the material. Hard magnets dissipate more energy during a hysteresis loop, thus requiring more energy to align the magnetization with the magnetic field. This explains why hard magnets make efficient memory systems. Energy dissipation from soft magnets is much less, which is why they are used in electronic systems.  In rate-dependent hysteresis, behavior similar to hard and soft magnets is observable by tuning the relaxation time. The area of hysteresis loop increases linearly with the relaxation time in the strong dissipation regime and decreases inversely linearly with the relaxation time in the weak dissipation regime. In the middle where the relaxation time is comparable to the period of driving, the dissipated energy reaches a maximum. This universal behavior will be clearly demonstrated in magnetic and cold-atom systems. Moreover, we found that the response function of rate-dependent hysteresis resembles the response function of a damped oscillator in the massless limit. Interestingly, similar examples can be seen in Kramers transition-rate theory of chemical reactions \cite{Kramer_Rev} and thermal transport in a classical 1D lattice \cite{Lepri03,Thermal_Trans}. 

While the framework of rate-dependent hysteresis analyzed here can be generic, interactions may invalidate the framework. However, there exist systems with interactions exhibiting rate-dependent hysteresis. The Landau-Lifshitz-Gilbert model of magnetization is an example \cite{Liu16}. Here we will show that, in cold-atom systems, artificial spin-orbit coupling \cite{SOC_YJLin,Wang12,Cheuk12} also permits rate-dependent hysteresis.  Moreover, a spin-orbit coupled Fermi gas will be shown to exhibit the universal behavior of dissipated energy.

This paper is organized as follows. Two examples of rate-dependent hysteresis in noninteracting systems, a two-state paramagnet and ultracold fermions in a ring, are analyzed in Sec.~\ref{sec:nonint}. Both systems exhibit hysteresis loops in the presence of tunable dissipation and periodic driving. The dissipated energy as a function of the relaxation time exhibits universal behavior and is summarized in Sec.~\ref{sec:dissipation} with detailed explanations. Sec.~\ref{sec:int} addresses how interactions can invalidate the framework implemented here and shows that spin-orbit coupled Fermi gases still follows the framework and exhibits rate-dependent hysteresis. Sec.~\ref{sec:conclusion} concludes our work.

\section{Hysteresis in noninteracting systems with dissipation}\label{sec:nonint}
The principle behind rate-dependent hysteresis, arising from relaxation towards equilibrium, can be demonstrated in simple noninteracting systems. We begin with a two-state paramagnet subjected to a periodically modulated magnetic field, and then present an extension following the same formalism for ultra-cold fermions in a ring-shape trap driven by a periodically modulated artificial vector potential. 

\begin{figure}
\includegraphics[width=3.4in]{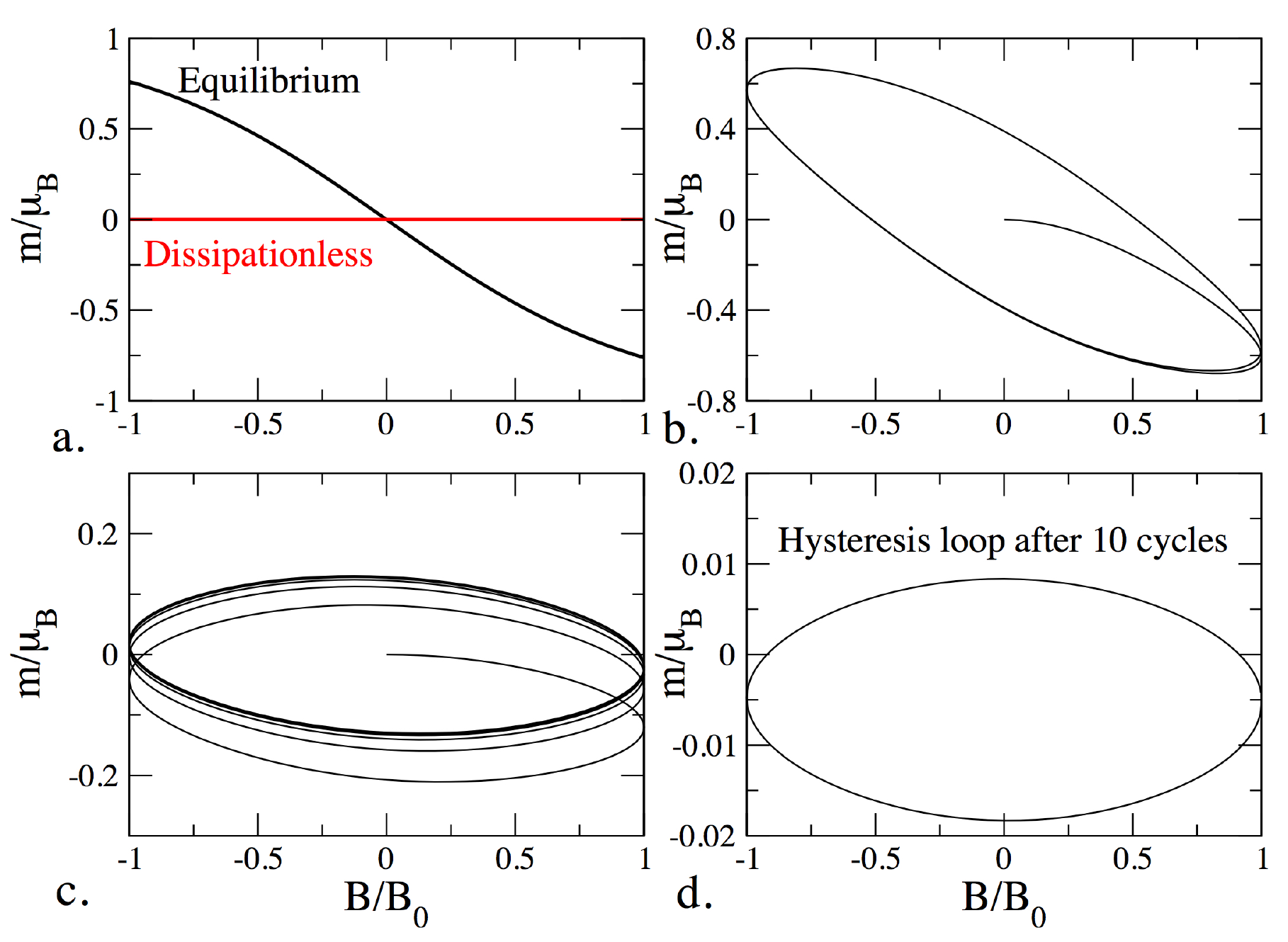}
\caption{Magnetization $m$ of a two-state paramagnetic system driven by an oscillating magnetic field $B$. (a) The equilibrium (black) and dissipationless (red) magnetization for the two-state paramagnet as a function of the magnetic field. Hysteresis loops of the magnetization emerge as the magnetic field  is periodically modulated with dissipation characterized by a relaxation time of (b) $\tau/t_p = 0.1$, (c) $\tau/t_p = 1$, (d) and $\tau/t_p = 10$. Here $t_p$ and $\tau$ are the period of the magnetic field and the relaxation time, respectively. We choose $\beta\mu_B B_0=1$, where $\beta=1/(k_B T)$, $\mu_B$ is the magnetic moment, and $B_0$ is the magnitude of the magnetic field. }
\label{Fig:Hyst_TS}
\end{figure}

\subsection{Two-State paramagnet}
In a simplified model, a two-state paramagnet has two energy levels labeled by the spin in the presence of a magnetic field, say in the $\hat z$ direction with magnitude $B$ \cite{Schro-ThermIntro}. The energy levels are  $E_{\sigma=\uparrow,\downarrow} = \mp\mu_B B$, where  $\mu_B$ is the magnetic moment of one spin. 
In equilibrium, the probability of seeing the spin-up or spin-down state can be calculated from the partition function. We define $\beta = 1/(k_B T)$, where $k_B$ and $T$ are Boltzmann constant and temperature, respectively. Explicitly,
$P_{\sigma} = e^{-\beta E_{\sigma}}/Z$ and $Z=2\cosh(\beta \mu_B B)$.
The magnetization is $m=\mu_B(P_{\uparrow} - P_{\downarrow})$, and in equilibrium 
\begin{eqnarray}\label{eq:meq}
m_{eq} = \mu_B\tanh(\beta \mu_B B).
\end{eqnarray}

In equilibrium, the magnetization as a function of the magnetic field follows the black curve in Figure~\ref{Fig:Hyst_TS}(a). In contrast, if there is no dissipation to relax the system, the distributions of the two states will remain the same as the initial condition. For example, if we consider a case with $B=0$ and $P_{\sigma=\uparrow,\downarrow}=1/2$ as the initial condition, then $m$ remains zero even when $B$ is turned on because the distribution cannot relax. The dissipationless case is shown as the red curve in Fig.~\ref{Fig:Hyst_TS}(a). We may consider the fully equilibrium case and the dissipationless case as the two extremes, and discuss general cases with finite dissipation. A simple and widely used formalism for modeling relaxation towards equilibrium is the relaxation approximation  
\begin{equation}
\frac{dP_\sigma}{dt} = -\frac{P_{\sigma}(t)-P_{\sigma,eq}(t)}{\tau}.
\end{equation}
Here $\tau$ is the relaxation time and $P_{\sigma,eq}(t)$ is the equilibrium distribution at time $t$. In a semiclassical picture, the relaxation time is the duration between scattering events and determines the characteristic time for the system to return to equilibrium. Relaxation of nuclear moments \cite{NuclearInduction} and electron properties in metals \cite{Ash_Merm} have been determined with the relaxation approximation and it proves to be versatile. 

Here we consider the system continually exposed to a periodic magnetic field $B=B_0\sin(\omega t)$ with $m(t=0)=0$. 
Time-evolution of the magnetization is calculated from the first-order differential equation describing the relaxation approximation:
\begin{equation}\label{eq:mrelax}
\frac{dm(t)}{dt}=-\frac{m(t)-m_{eq}(t)}{\tau}.
\end{equation} 
The explicit solution is
\begin{equation}\label{eq:mt}
m(t) = \frac{1}{\tau}\int_{0}^{t}e^{-(t-t')/\tau}\mu_B\tanh(\beta \mu_B B(t'))dt'.
 \end{equation}
 The time-dependent magnetization is shown in  Figure~\ref{Fig:Hyst_TS}(b)-(d).

The evolution of magnetization of the two-state paramagnet in the presence of an oscillating magnetic field clearly exhibits hysteresis. Moreover, after a few training cycles, hysteresis loops are formed. Using the oscillation period $t_p = 2\pi/\omega$ as the time unit, the magnetization from selected relaxation times $\tau/t_p = 0.1$, $1$, $10$ are plotted against the magnetic field over ten periods in Figure~\ref{Fig:Hyst_TS}(b-d). Training cycles are required for the initial condition to decay away, and the number of training cycles depends on $\tau/t_p$. When $\tau/t_p$ is small (large), the initial condition decays faster (slower) and less (more) training cycles are needed.
 Training cycles have been seen in the micro-states of artificial spin ice when subjected to a periodic magnetic field and self organization of nanoparticles undergoing periodic shearing \cite{SpinIce_Mem, SelfOrg_Nano}. 
 
The formation of the hysteresis loops can be understood in the weakly and strongly dissipative regimes as follows. The integrand of Eq.~\eqref{eq:mt} is a product of the exponential factor $\exp(-(t_f-t^\prime)/\tau)$ and a periodic function with period $t_p$. In the strongly dissipative regime $\tau/t_p\ll 1$, and the exponential factor severely suppresses the integrands if $t_f-t^\prime$ is finite. Therefore, only the contribution from the last cycle is visible and forms the hysteresis loop. In the other limit when $\tau/t_p\gg 1$, the exponential factor is basically a constant over many periods, and the result is a summation of many repeating cycles, which also produces hysteresis loops. Importantly, after proper training cycles, hysteresis loops can be observed for finite values of $\tau$ as shown in Figure~\ref{Fig:Hyst_TS}.

\subsection{Atomic current from noninteracting fermions in a ring}
The same theoretical framework can be applied to 
non-interacting fermions in a ring trap threaded with effective magnetic flux. Ultracold atoms are particularly suitable for realizing such a system, and we consider fermionic atoms trapped in a ring-shaped potential and driven by an artificial vector potential equivalent to an artificial magnetic flux. The setup is similar to Ref.~\cite{SF_Hysteresis} except here a current is driven by the artificial magnetic flux rather than a laser barrier. 
In absence of self interactions, the wave function of non-interacting fermions can be described in the single particle picture \cite{Landauer_Ring_PC}. 
The Hamiltonian describing noninteracting fermions in a ring subjected to an effective vector potential $A$ along the ring is
\begin{equation}\label{eq:HA}
H = \frac{1}{2m_f}(p_x + A)^2.
\end{equation}
Here $m_f$ is the mass of the atom, $x$ is the coordinate along the ring and is periodic in the circumference $L$ of the ring, the momentum operator is $p_x = -i\hbar \frac{\partial}{\partial x}$, and the dimensionless magnetic flux $\phi$ is derived from the artificial vector potential $A$ circulating along the ring via $\Phi = AL = 2\pi \hbar \phi$.

The energy spectrum is obtained by solving the Schr\"odinger equation with the eigenfunction 
\begin{equation}
\psi_n =  \sqrt{\frac{1}{L}}e^{ik_n x}.
\end{equation}
Here $k_n = 2\pi n/L$ with $n$ labeling the states. Importantly, this wave function is the exact eigenfunction for any value of the vector potential, as one can check by applying the Hamiltonian~\eqref{eq:HA} on it. The energy eigenvalues, however, explicitly depend on $A$ and have the following expresion 
\begin{equation}\label{eq:En}
E_n = E_u(n+\phi)^2.
\end{equation}
Here $E_u=(2\pi\hbar)^2/(2m_fL^2)$ serves as the unit of energy in this case. By plotting $E_n$ of several $n$ values as a function of $\phi$, one can see that the energy spectrum repeats itself as $\phi$ changes to $\phi+1$ \cite{MesoPhys_Akkermans}. Following the idea of Brillouin zone in solid-state physics, we may focus on the "first Brillouin zone" of the spectrum with $-0.5 < \phi < 0.5$.  Moreover, the states with $\pm n$ are degenerate at $\phi = 0$ and the degeneracy is lifted when $\phi>0$.  When $\phi = 0$, the generalized Bloch theorem rules out a finite equilibrium current in the absence of magnetic flux \cite{GenBlochTh}.

The current can be obtained from the conservation of particles
$\partial \rho/\partial t = -\nabla \cdot j$.
Since the particle density is $\rho = \sum_{n}\psi_n^* \psi_n$, the current from each eigenstate can be found from $j_n = (\hbar/i2m_f)(\psi_n^*\nabla \psi_n - \psi_n \nabla \psi_n^*)$. Equivalently, the current can be found by \cite{MesoPhys_Akkermans} 
\begin{eqnarray}
j_n &=& -\frac{\partial E_n}{\partial \Phi} = J_0(n+\phi).
\end{eqnarray}
Here $J_0=(2\pi\hbar)/(m_f L^2)$. 
At zero temperature, the total current can be found by summing the currents from the state below the Fermi energy. The expression can be generalized to finite temperatures by including the distribution functions of the energy states and is given by 
\begin{equation}\label{eq:Jtot}
J_{Tot} = \sum_n j_n f_n.
\end{equation}
Here $f_n$ denotes the population of the state $\psi_n$.
Thus the density distribution affects the total current, and a relaxation process is necessary to induce re-distribution of the density distribution, which will shift the current aways from the dissipationless limit. 
In thermal equilibrium the distribution should be the Fermi-Dirac distribution 
\begin{equation}\label{eq:fneq}
f_{n,eq} = {1 \over e^{\beta(E_n - \mu) + 1}}.
\end{equation}

\begin{figure}
\includegraphics[width=3.4in]{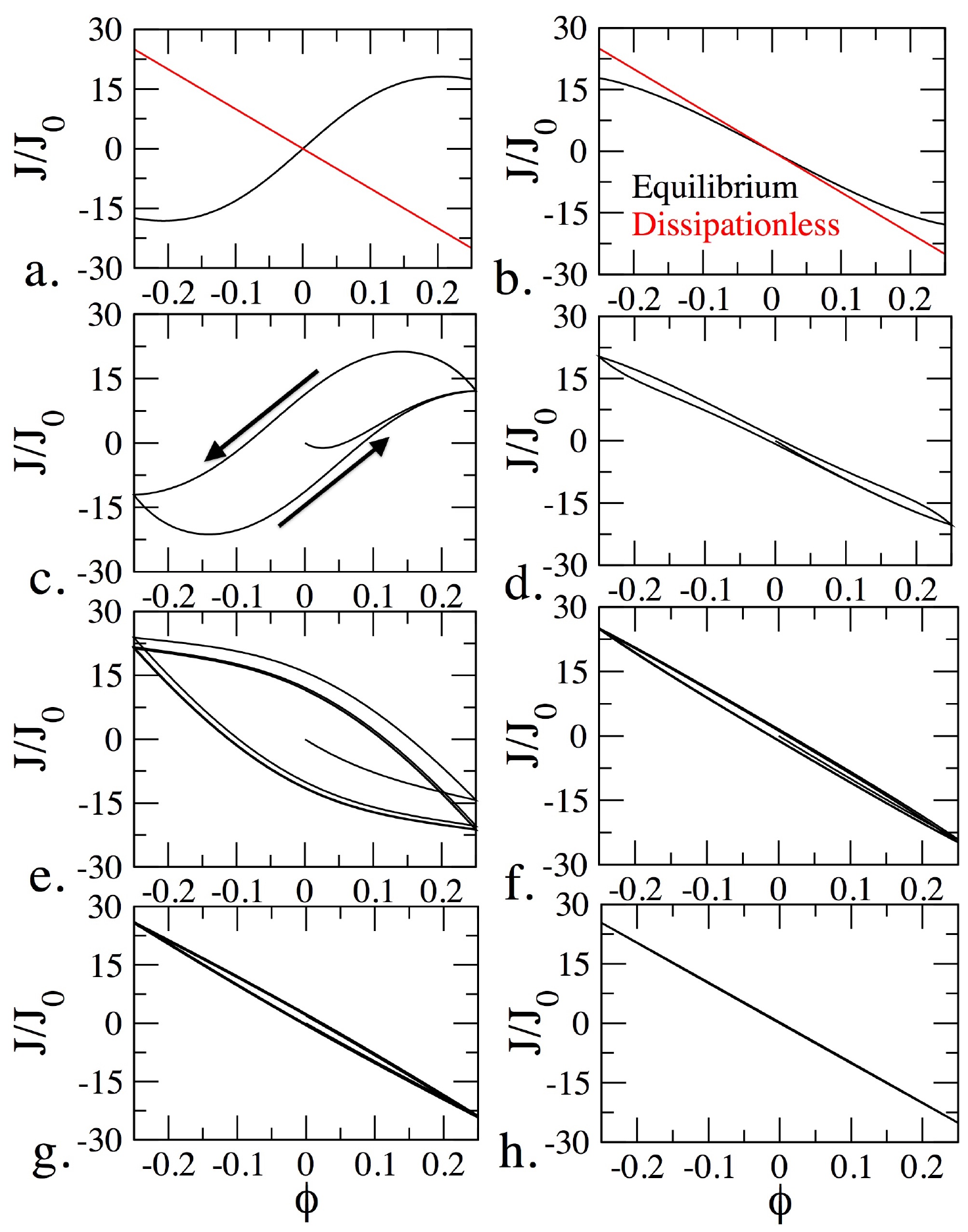}
\caption{Total atomic current $J_{Tot}$ over the flux $\phi$ at $k_B T/E_u = 10$ for noninteracting fermions in a ring. The left (right) column shows a system with $N_{tot}=100$ ($N_{tot} = 101$)  particles. (a) and (b): The equilibrium (black) and dissipationless (red) currents. (c-h) The current exhibit hysteresis loops as the flux is periodically modulated in the presence of dissipation with a relaxation time of (c,d) $\tau = 0.1t_0$, (e,f) $\tau = t_0$, (g,h) and $\tau = 10t_0$. }
\label{Fig:Hyst_Curve}
\end{figure}

Persistent current refers to the current on a ring induced by a magnetic flux through the ring when the system is in thermal equilibrium \cite {MesoPhys_Akkermans}. The persistent current thus corresponds to the equilibrium current shown in Fig.~\ref{Fig:Hyst_Curve} (a). For each value of $\phi$, the corresponding equilibrium distribution $f_{n,eq}$ should be used to evaluate the persistent current. Summing over all energy levels to determine total equilibrium current results in a dependence on the total particle number as shown in  Figure~\ref{Fig:Hyst_Curve}(a,b). At finite temperatures, the persistent current shows smooth dependence on $\phi$. When the flux is absent  ($\phi=0$), there will be no genuine persistent current in the system and this is consistent with the generalized Bloch theorem \cite{GenBlochTh}.

Maintaining an equilibrium current as the flux varies, however, requires strong dissipation. Imagine that the initial state is in equilibrium with no effective magnetic flux, and then a flux is turned on. In the noninteracting fermion case, each eigenstate remains the same eigenstate, but the corresponding eigenvalue shifts. In absence of any dissipation, the population of each eigenstate will not re-distribute because the Hamiltonian does not have cross-correlation between different states, and changing the population distribution requires an external mechanism to reshuffle the particles at different eigenstates. In a system of noninteracting fermionic atoms, there is no dissipation because of a lack of scattering mechanism. Therefore, observations of genuine persistent currents by ramping up the effective vector potential can be a challenging task.  First, the system can not reach thermal equilibrium after the vector potential changes, so a non-equilibrium current will be observed instead. Second, the current can be small because when $j_n$ is summed over all levels weighted by the distribution, the contributions from $n$ and $-n$ mostly cancel. Realizing a fermionic persistent current in a cold atom system thus requires some form of relaxation by introducing external dissipation mechanisms and also accurate measurements of the current.

In stark contrast, for electrons in metallic systems dissipation is strong and inevitable because scatterings result from crystal defects and impurities. As shown in the two-state paramagnet example, a widely implemented formalism for describing the relaxation of a system subjected to a dissipative environment is the relaxation approximation. Here the relaxation approximation for an $n$-level system governed by the Fermi-Dirac statistics leads to 
\begin{equation}\label{eq:dfn}
\frac{\partial f_n}{\partial t} = {-(f_n-f_{n,eq}) \over \tau},
\end{equation}
where $f_{n,eq}$ is the equilibrium distribution function for the $n$-th state and we assume the relaxation time $\tau$ is the same for all states \cite{Ash_Merm}. The relaxation time estimates the duration between two consecutive scattering events of a particle. For an electron in typical metals, $\tau$ is roughly $10^{-14}$s \cite{Copper_Relax, Gold-Relax, Silver_Relax}. The fast relaxation for electrons in metals allows for direct measurements of persistent current \cite{PC_Copper, PC_GoldLoop}. On the other hand, in the natural setting ultracold atoms are dissipationless, which hinders the emergence of fully equilibrium persistent current. Nevertheless, dissipation may be introduced in cold-atom systems by allowing the system to interact with background atoms or photons. Moreover, by engineering the dissipation to tune the relaxation time, nonequlibrium currents will reveal the competition between the driving vector potential and the dissipation. The collision rate of cold fermions in a boson bath is tunable by varying the density of bosons \cite{CollisionTransport}, and the investigation of different relaxation times leads to dynamical hysteresis effects. 

The solution to Eq.~\eqref{eq:dfn} at time $t_f$ is
\begin{equation}\label{eq:f_n}
f_n(t_f) = e^{-t_f/\tau}f_n(t=0) +  \int_0^{t_f}e^{-(t_f-t)/\tau}\frac{f_{n,eq}(t)}{\tau} dt.
\end{equation}
Here the equilibrium distribution at time $t$ is given by Eq.~\eqref{eq:fneq} with $\phi(t)$ in the energy dispersion $E_n(t)$ from Eq.~\eqref{eq:En}. The equilibrium chemical potential $\mu(t)$ at time $t$ for a given $\phi(t)$ should conserve the total particle number $N_{tot}$ and can be determined by the relation 
\begin{equation}
N_{tot} = \sum_n{1 \over e^{\beta(E_n(t) - \mu(t))}+1}.
\end{equation}
 In our simulations we choose $\phi(t)$ to be periodic with period $2t_0$ and a linear dependence on the time. Therefore, $\phi(t)$ shows a saw-tooth pattern when plotted as a function of $t$. As we will show, the system exhibits rate-dependent hysteresis, so the choice of the periodic form of $\phi(t)$ is not important. We also choose $\phi(t)$ to oscillate between $-0.25$ and $0.25$. 
 The unit of time will be $t_0$, so the relaxation time is expressed as $\tau/t_0$. For a given $\tau$ and $t_f$, the corresponding $\phi(t_f)$ is determined. Then $E_n(t_f)$, $\mu(t_f)$, and $f_{n,eq}(t_f)$ can be evaluated. The total  current $J_{Tot}(t_f)$ is then given by Eq.~\eqref{eq:Jtot}. This procedure produces a pair of $(\phi(t_f), J_{Tot}(t_f))$, and by sweeping $t_f$ we traced out the dynamic evolution of the current in the presence of dissipation. 

A competition between the driving period and the relaxation time is necessary to observe hysteresis in a cold atom system. In the limits when the system is always brought into equilibrium immediately ($\tau/t_0\rightarrow 0$) or in the absence of dissipation ($\tau/t_0\rightarrow \infty$), the current will simply follow the fully-equilibrium curve or the dissipationless current in Figure~\ref{Fig:Hyst_Curve}(a,b) without hysteresis. In the presence of finite dissipation, the current will lie somewhere between the equilibrium and dissipationless current. After a couple of  training cycles, the current clearly exhibits hysteresis loops as shown in Figure~\ref{Fig:Hyst_Curve} for selected values of $\tau$ and numbers of fermions $N$. The larger the relaxation time $\tau$ is, the more training cycles are needed to form hysteresis loops. Moreover, the size of hysteresis depends explicitly on the relaxation time $\tau$, and we will analyze the dependence later.

Therefore, starting with noninteracting fermions with dissipation, hysteresis loops of mass current are expected to emerge. This is in contrast to previous work \cite{SF_Hysteresis} where atomic superfluids with interactions are used to demonstrate hysteresis loops. By analogue with rate-dependent hysteresis in a paramagnet \cite{MagHyst}, hysteresis loops can arise without the need of self interactions. Therefore, cold-atoms are particularly suitable for demonstrating hysteresis behavior with and without interactions. Dissipation of the noninteracting fermions is from inter-species interactions with a background of atoms of a different species or electromagnetic waves. The inter-species interactions may induce higher-order corrections to the fermions acting like effective self-interactions. However, the induced interaction is expected to be weak and considered as secondary effects.

\subsection{Experimental Realization of Hysteresis in Noninteracting Fermions}
Measurement of atomic persistent current is achievable with current technology, and has already been facilitated for bosons in the superfluid phase using a rotating laser barrier to stir the atoms \cite{SF_Hysteresis}.  The ring trap can be constructed by using Laguerre-Gaussian (LG) beams for azimuthal confinement and a pair of laser sheets for tight vertical confinement \cite{Ring_Lattice, SF_Qubit_Amico}. Atoms propagating in a ring-shape trap with periodic boundary conditions could further allow cold-atom studies of entanglement between boson species \cite{AG_Entangle} and persistent current of interacting  bosons\cite{Minguzzi_BosonPC}. 

There have been a variety of methods to induce artificial gauge fields, which is done by finding methods mimicking the momentum shift due to the Lorentzian force \cite{AG_Theory, Lin_MagGauge, ColdAtom-SOC-Rev}.  In cold atoms, atom-light coupling from an external laser field causes an effective vector potential and corresponding magnetic field. Use of gauge fields from the coupling of the atoms and laser field is an advantageous tool, and has lead to measurements of new topological phases not easily realizable in condensed matter systems \cite{Triangle_SpinIsing, Shaken_Topology}.

To mimic a circulating vector potential along the ring, an Abelian gauge field is sufficient to induce the required effective magnetic flux. The direction of the gauge field is along the direction of lightwave momentum vector, so the light must propagate along the direction of the ring for a flux to be present. The optical ring potential is confined in the xy-plane meaning the lightwave momentum vector must propagate along the circular ring potential in the 2D-plane. The LG beam will serve as an appropriate tool to obtain a circular momentum flux \cite{Gauge_LGbeam}. Modes of the beam can be selected from the angular momentum $l$ in units of $\hbar$, and the value of $l$ determines the accumulated phase as the atoms traverse the ring. If $l = 0$, no phase is accumulated which is equivalent to having no effective magnetic field. A beam with $l = 1$ acts as an Abelian gauge field, which is the type needed to drive the current, because the atoms  accumulates a phase of $\pi$ after circulating the ring once. A non-Abelian gauge field can be selected when $l=2$, but such a complicated field is not needed for the system in question.

The above methods address how to trap the atoms and create a synthetic magnetic flux, but the key component is a background dissipation mechanism. To bring the system towards thermal equilibrium, the fermions need to interact with the environment. In condensed matter systems, scattering of particles with impurities occurs on a fast timescale, allowing for equilibrium current to be observed. It is necessary to add a form of scattering to the particles trapped in the ring potential. 

Introducing dissipation through Hamiltonian engineering has been highly useful for creating pure entangled states, and a variety of methods  to couple the system to its environment have been explored using electron beams and the release of a Bogoliubov quasi-particle after undergoing a Raman transition to an excited state \cite{Ott_DissEng, Diss_QuantumWire, DissHamiltonianEng}. However, a simpler and more direct method utilizes the broad tunability of atomic fermion-boson mixtures. By immersing the fermions trapped in a ring potential into a Bose-Einstein Condensate (BEC), collisions between the two species can facilitate the dissipative processes caused by impurities in metals. The relaxation time $\tau$ can be controlled by adjusting the number of bosons in the atomic cloud, and collisions occur on the millisecond timescale \cite{CollisionTransport}. 

Particle loss is an apparent concern when artificially invoking collisions because the collision process can excite a fermion out of the potential barrier and can affect the direct observation of the system. However, it has been experimentally observed that fermions experience very little loss while the depletion of bosons is prominent in an atomic fermion-boson mixture \cite{BF_Mix_3D}. The dynamics of a Bose-Fermi mixture is also in favor of experimental realization. It has been shown theoretically and experimentally the dissipative mechanism can induce a current of non-interacting fermions in a periodic potential by acting as a driving force for an initially insulating phase \cite{CollisionTransport, CollisionTransport_Theory}. In direct contrast,  localization is observed in the bosonic cloud when interacting with fermionic impurities \cite{Fermion_Impurities}. Moreover, advanced imaging techniques allow separate resolutions of the fermionic and bosonic quantities \cite{BF_Mix_3D, Quant-Microscope, Absorp-Image}, and here we will focus on the fermionic system by treating the bosons as the environment.

\begin{figure}
\includegraphics[width=3.4in]{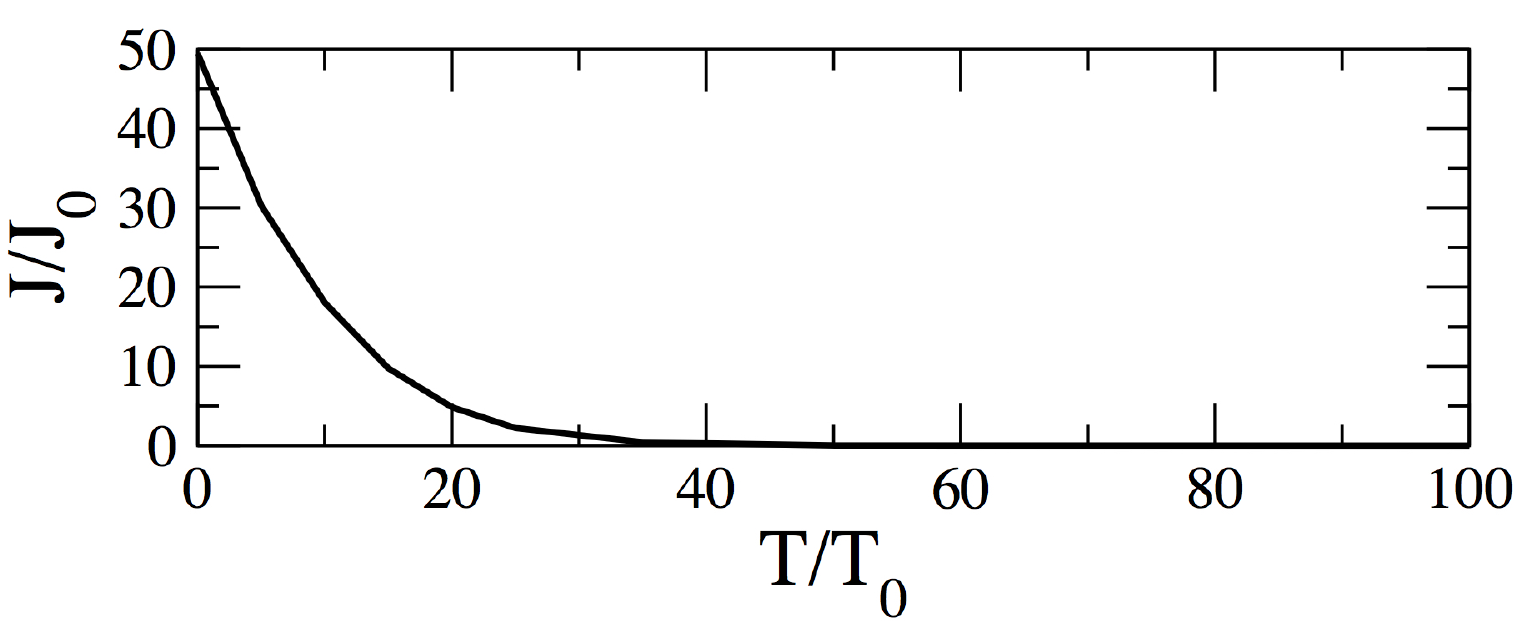} 
\caption{Maximal persistent current as a function of the normalized temperature $T/T_0$ for $N = 101$ noninteracting fermions.}
\label{Fig:Temp}
\end{figure}

The amplitude of the fully equilibrium current is highly sensitive to temperature, and the value decreases exponentially as the system gets warmer. Figure~\ref{Fig:Temp} shows the amplitude of the equilibrium current, $\sum_{n}j_n$, for $N_{tot}=101$. The Fermi temperature is $T_F = 2500T_0$, where $T_0=E_u/k_B$. In order to observe the current, the system may need to be cooled to $1\%$ of the Fermi energy. Using a compensated lattice technique the system can be compressed and cooled to nearly the ground state\cite{CompensatedLattice, FermionCompress}. A previous experiment using mesoscopic, copper rings also reported an exponential decrease of the electronic persistent current as temperature increases, but were able to measure the persistent current when the system was cooled to $80$ mK since copper has a Fermi temperature of about $8000$K \cite{PC_Copper, Ash_Merm}.

\section{Universal behavior of dissipated energy}\label{sec:dissipation}
The relaxation dissipates energy as the system is being driven. The system acts as a media for transferring the energy input from the driving force ($B$ in the two-state paramagnet or $\phi$ in the atomic current) to the reservoir responsible for dissipation. The amount of transferred energy is reflected by the amount of dissipated energy of the system, $\Delta E$. Importantly, $\Delta E$ can be measured by the area of the hysteresis curve.  By calculating the area of one hysteresis loop, how much energy is dissipated in a cycle can be determined. For the two-state paramagnet system, the energy dissipated in one hysteresis loop is $E = \oint m\cdot dB$ and for the atomic current in a ring is $E = \oint J_{Tot}\cdot d\Phi$. A smaller hysteresis loop means a cycle dissipates less energy. When $\tau$ is small (compared to the period of driving), the system relaxes into thermal equilibrium easily, so the current basically follows the driving and the energy transferred by the system in one direction (say, when $B$ or $\phi$ increases) is mostly canceled by the other half of the cycle. When $\tau$ is large, the system has a strong memory of its initial state because the relaxation time is too long. The current thus does not respond promptly to the driving, and again the energies from the two halves of a cycle tend to cancel. Prominent hysteresis loops are possible when $\tau$ is of the same order of the period of driving. In this case, the current lags behind the driving, so the system dissipates and absorbs different amounts of energy in one cycle. As a consequence, there is a net energy flow from the driving force to the reservoir, and its magnitude is proportional to the area of the hysteresis loop.

For the atomic current in a ring, the dissipated energy inherits the dependence of the current on the number of particles. Since the fully equilibrium current and fully dissipationless current limit the range of nonequilibrium current for finite dissipation, the size of hysteresis loop is constrained by the separation of the two limiting cases. For odd numbers of fermions, the fully equilibrium current is close to the dissipationless current according to the initial distribution, so the hysteresis loop is in general small. In contrast, for even numbers of fermions, the difference between the fully equilibrium and dissipationless currents are significant, so the hysteresis loop can have a larger area. Figure~\ref{Fig:Hyst_Curve} clearly demonstrates the dependence of the hysteresis loop on the fermion number.  

\begin{figure}
\includegraphics[width=3in]{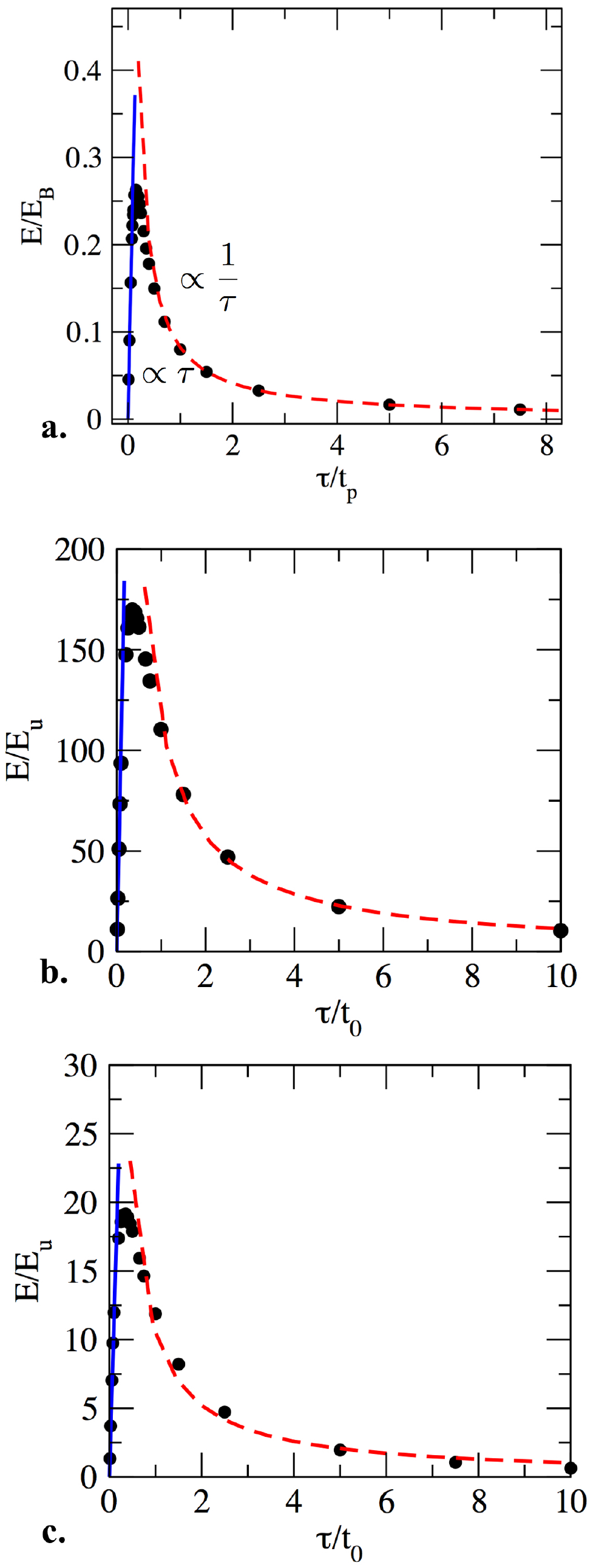}
\caption{Energy dissipated in one hysteresis loop as a function of the relaxation time $\tau$ for (a) the magnetization of the two-state paramagnet and (b), (c) the atomic currents from $N = 100$ (b) and $N = 101$ (c) noninteracting fermions in a ring. In the strongly dissipative regime, $E \propto \tau$ is apparent with small $\tau$. In the weakly dissipative regime, the inverse relationship $ E \propto 1/\tau$ with large $\tau$ is observable. The blue and red curves show the $\tau$ and $1/\tau$ dependence in the three cases. The energy units are $E_B=\mu_B B_0$ for (a) and $E_u=(2\pi\hbar)^2/(2m_f L)^2$ for (b) and (c). }
\label{Fig:Hyst_Area}
\end{figure}

The area of the hysteresis curve, which represents the energy transferred from the driving to the reservoir, has a non-monotonic dependence on the relaxation time. The area increases with a linear dependence on the relaxation time, $E \propto \tau$, but after reaching a critical value the area decreases inversely, $E \propto 1/\tau$ as shown in Figure~\ref{Fig:Hyst_Area}. The dissipated energy reaches a maximum when $\tau$ is of the same order of the driving period. This behavior can be explained through evaluation of the integrals in Eqs.~\eqref{eq:mt} and \eqref{eq:f_n}, but a closer evaluation demonstrates deep connections with a damped harmonic oscillator in the massless limit. We begin with the behavior in the small and large $\tau$ limits. After enough training cycles, the leading term from the initial condition is diminished. The relaxation of the distribution then takes the form
\begin{equation}
g(t) = \frac{1}{\tau}\int_{0}^{t}e^{-(t-t')/\tau}g_{eq}(t') dt'.
\end{equation}
When $\tau >> t_p$, the value of the exponential is roughly constant during one cycle, so
$g(t) \approx \frac{1}{\tau}\int_{t-t_p}^{t}g_{eq}(t') dt' \propto \frac{1}{\tau}$. Here $t_p$ is the period of the driving. 
Thus in this regime the area decreases inversely with the relaxation time. In the regime where $\tau << t_p$, a small amount of $t-t'$ leads to a severe suppression due to the exponential factor. Thus, we may focus on a small period $\epsilon(\tau)$ prior to the final time $t$. The major contribution to the integral is from this small period and the equilibrium distribution function can be considered constant over the small period in time.
$g(t) \approx \frac{g_{eq}(t)}{\tau}\int_{t-\epsilon(\tau)}^{t}e^{-(t-t')/\tau} dt'$
By setting $\tilde{t} = t - t'$ the integral becomes
\begin{eqnarray}
g(t) &\approx& \frac{g_{eq}(t)}{\tau}\int_{0}^{\epsilon(\tau)}e^{-\tilde{t}/\tau} d\tilde{t} = g_{eq}(t)(1-e^{-\epsilon(\tau)/\tau}).
\end{eqnarray}
Expanding $\epsilon(\tau) = a\tau + b\tau^2 + ...$ leads to $1- e^{-(a+b\tau)} \approx 1-e^{-a}( 1- b\tau)$. Only first-order terms are considered with small $\tau$, so $g(t) \approx g_{eq}(t)(1-e^{-a} +e^{-a} b\tau)$. Taking the limit $\lim_{\tau \rightarrow 0}g(t) = 0$ forces $a = 0$, leaving the linear dependence on the relaxation time $\tau$.

A fully analytic analysis of the area of hysteresis loop in rate-dependent hysteresis of the two-state paramagnet case in the full range of $\tau$ can be performed following the discussion in Ref.~\cite{MagHyst}. The driven magnetization in the relaxation approximation is described by Eq.~\eqref{eq:mrelax}
The magnetization $m_{eq}$ in equilibrium is given by Eq.~\eqref{eq:meq}, and the evolution of $m(t)$ is given by Eq.~\eqref{eq:mt}.
 In the high temperature limit, $m_{eq} \approx m_0 B(t)$ with $m_0=\beta \mu_B^2$. Then,  
\begin{equation}
m(t) \approx \frac{m_0}{\tau}\int_{-\infty}^{t}e^{-(t-t')/\tau}B(t')dt'
\end{equation}
The magnetic field $B(t)$ acts as a driving force. The magnetic susceptibility $\chi(\omega)$ is the response function connecting magnetization with magnetic field \cite{MagHyst} in the frequency domain.
\begin{eqnarray}
m(\omega) &=& \chi(\omega)B(\omega)
\end{eqnarray}
Explicitly,
\begin{equation}\label{eq:chi}
\chi(\omega) = \frac{\chi_0}{1-i\omega \tau}.
\end{equation}
Here $\chi_0=m_0$.

A damped harmonic oscillator is described by the Langevin equation $ M\ddot{x} +kx+\gamma \dot{x} = \eta$, where $M$, $k$, and $\gamma$ denote the mass, spring constant, and friction coefficient, and $\eta$ is a random force with white-noise spectrum \cite{ChaikinCMBook}. When the oscillator is overdamped, $\gamma \gg M\sqrt{M/k}$, the inertial term in the equation can be ignored and the response function has exactly the same structure as Eq.~\eqref{eq:chi}. In the magnetization case, however, we are not taking the overdamping limit. Instead, the system corresponds to the massless limit of the Langevin equation, and the response function \eqref{eq:chi} applies to any value of $\tau\propto 1/\gamma$. Importantly, the energy dissipated in one hysteresis loop is $E=\pi B_0^2 \chi''(\omega)$, where $\chi''(\omega)$ is the imaginary part of the response function. Explicitly, 
\begin{eqnarray}
E &=& \pi m_0B_0^2\frac{\omega \tau}{1+\omega^2 \tau^2}.
\end{eqnarray}
Here $\omega=2\pi/t_p$ with $t_p$ denoting the period of the periodic driving. Therefore, $E\propto \tau$ when $\tau \ll t_p$ and $E\propto 1/\tau$ when $\tau \gg t_p$.

Interestingly, the linear and inverse-linear dependence on dissipation have been demonstrated elsewhere in chemistry and thermal physics. The Kramers transition rate theory models a chemical reaction rate by the Brownian motion of a particle moving out of a metastable well \cite{Kramer_Rev}. Due to the fluctuation-dissipation theorem, the magnitude of the random force in the Langevin equation increases with the friction coefficient. When the friction is small, there are insufficient kicks to move the particle out of the well, and the transition rate increases linearly as the friction coefficient increases. After the crossover point, the friction becomes large. Then, the particle has more chance of climbing up the energy barrier, but the violent kicks randomly throw the particle back to the well. In the strong friction regime, the transition rate is suppressed by the inverse of the friction coefficient. By plotting the transition rate as a function of the friction coefficient, there is a linear regime and an inverse-linear regime connected by a maximal transition rate in between.

Similar effects are also observable when two reservoirs with different temperatures are connected by a 1D classical lattice \cite{Lepri03,Thermal_Trans}. The thermal conductance exhibits similar dependence on the friction coefficient coupling the lattice to the reservoirs. When the friction is small, the heat conduction is limited by how much energy is pumped into the lattice from the reservoir. Increasing the friction allows for higher energy input and increases thermal conduction linearly. When the friction is strong, however, the lattice sites coupled to the reservoirs are basically in thermal equilibrium with the reservoirs and exhibit different energy spectra from the rest of the lattice. The mismatch of the energy spectra suppresses heat conduction. As a consequence, heat conduction decreases with the inverse of the friction coefficients. A maximal thermal conductance emerges when the two regimes meet in between.

\section{Interaction Effects}\label{sec:int}
So far we only present rate-dependent hysteresis in noninteracting systems. One important feature of the framework built here is that the eigenstates remain the same regardless of the periodic driving force. The eigenvalues, nevertheless, depend explicitly on the driving. By examining the two-state paramagnet and the Fermi gas in a ring, the following structure emerges 
\begin{equation}
H(R)\ket{\psi_n} = E_n(R)\ket{\psi_n}.
\end{equation}
Here $R$ denotes the driving force, which corresponds to $B$ in the two-state paramagnet or $\phi$ in the atomic current in a ring. Since the basis $\ket{\psi_n}$ is independent of $R$, we can focus on their population distribution. Relaxation mechanisms are then introduced to redistribute the population in the eigenstates.

This framework can be invalidated in the presence of interactions, which usually  correlate the eigenstates of the noninteracting Hamiltonian and make them explicitly depend on the driving force. Then it is no longer useful to analyze the population on the eigenstates at a given instance because the eigenstates at a different instance will be different. In the following we give some illustrations of how interactions invalidate the hysteresis mechanism presented here. Interestingly, there are special cases where selected types of interactions still allow the framework to describe rate-dependent hysteresis and we will summarize some examples, too.

\subsection{Limitation of the relaxation framework}
For the two-state paramagnet we assume the spins are only influenced by a modulating magnetic field in the $z$-direction, $B_z(t)=B_0\sin(\omega t)$. This will allow the $z$-directional eigenstates to be invariant under the drive. A simple modification of the problem is to add a constant transverse magnetic field and break the invariance. The corresponding Hamiltonian is 
\begin{equation}
 H = B_x\hat{\sigma}_x   + B_z(t) \hat{\sigma}_z.
 \end{equation}
The eigenstates at a given time $t$ is
\begin{eqnarray}
\ket{n,t}_{+} &=& \frac{1}{N}
\begin{pmatrix}
-B_x\\
B_z(t) - \sqrt{B_z(t)^2 +B_x^2}\\
\end{pmatrix} \nonumber \\
\ket{n,t}_- &=& \frac{1}{N}
\begin{pmatrix}
-B_z(t) + \sqrt{B_z(t)^2 +B_x^2}\\
-B_x\\
\end{pmatrix}.
\end{eqnarray}
Here $N=\sqrt{B_x^2+(B_z(t)-\sqrt{B_z(t)^2 +B_x^2})^2}$ being a normalization factor, and the corresponding eigen energies are 
$E_{\pm} = \pm\sqrt{B_z(t)^2+B_x^2}$.

As time evolves, the magnetic field $B_z(t)$ at each instance introduces a different set of eigenstates, and $\bracket{n,t_1}{n,t_2} \neq 1$. With each increment of time, not only the eigenvalues but the eigenstates are different from the previous one. In such a system, it is more advantageous to choose a fixed basis (such as the eigenstates in the absence of interactions) and analyze the evolution of the density matrix instead.

For multi-component ultracold fermions in a ring, contact interactions between different components due to collisions invalidate the single-particle picture. To find the ground state, one has to diagonalize the full Hamiltonian taking into account correlations between particles. For a simplified delta-function  interaction between two particles modeling the contact interactions, one has to resort to the Bethe Ansatz for finding the ground state and its energy. Even for only two particles in a ring subjected to an artificial vector potential, one has to match the boundary conditions of the wave functions. The consistent wavefunction then requires the wave vectors to be solved from coupled equations \cite{BA_Impurity}. Adjustment of the momentum of one particle alters the other, so one cannot fill the two particles in an arbitrary way like the noninteracting case. The complexity of solving the Bethe Ansatz increases rapidly as more particles are added to the system, so it will be a great challenge to verify whether rate-dependent hysteresis still survives in the presence of strong interactions. It is clear, though, the framework for noninteracting fermions does not apply to systems with contact interactions.

Hysteresis loops, however, can be found in some interacting systems. For example, the Landau-Lifshitz-Gilber (LLG) model describes spin injection from a fixed layer into a free layer (see Refs.~\cite{IntroSpintronics,KamenevBook} for a review), and effective interactions and dissipation are incorporated. By driving the LLG model with an oscillating magnetic field, hysteresis loops of magnetization are found in simulations \cite{Liu16}. For atomic currents in ring-shape potentials, here we show that hysteresis loops are observable in the presence of Rashba spin-orbit coupling.

\subsection{Hysteresis in Rashba Spin-Orbit Coupled Fermi Gases}
Here we present another case where rate-dependent hysteresis can be found in interacting systems.  Although the spin-orbit couplings in cold-atom experiments can be more general \cite{SOC_YJLin,Wang12,Cheuk12}, here 
we consider two-component ultracold fermions with artificial Rashba spin-orbit coupling (RSOC) \cite{LaserSOC_Spielman, NAGauge-Darkstate, NAGauge-Lewenstein, SpinHall_Gauge, AG_Colloq, Rashba_Rev, CA_SOC_Theory}. Coupling of electronic spin to its momentum arises from an effective field produced by the moving electron in the presence of Coulomb fields \cite{Rashba_orig, Rashba_OscEffects, NanowireSOC, SOC_spinfilter}.  Although cold-atoms are charge neutral, spin-orbit interactions can be engineered with similar experimental techniques as artificial gauge fields \cite{SOC_YJLin,Wang12,Cheuk12}. Coupling is induced in dressed atomic states by a pair of Raman lasers and the coupling strength is dependent on the momentum of the laser. 

\begin{figure}
\includegraphics[width=2.4in]{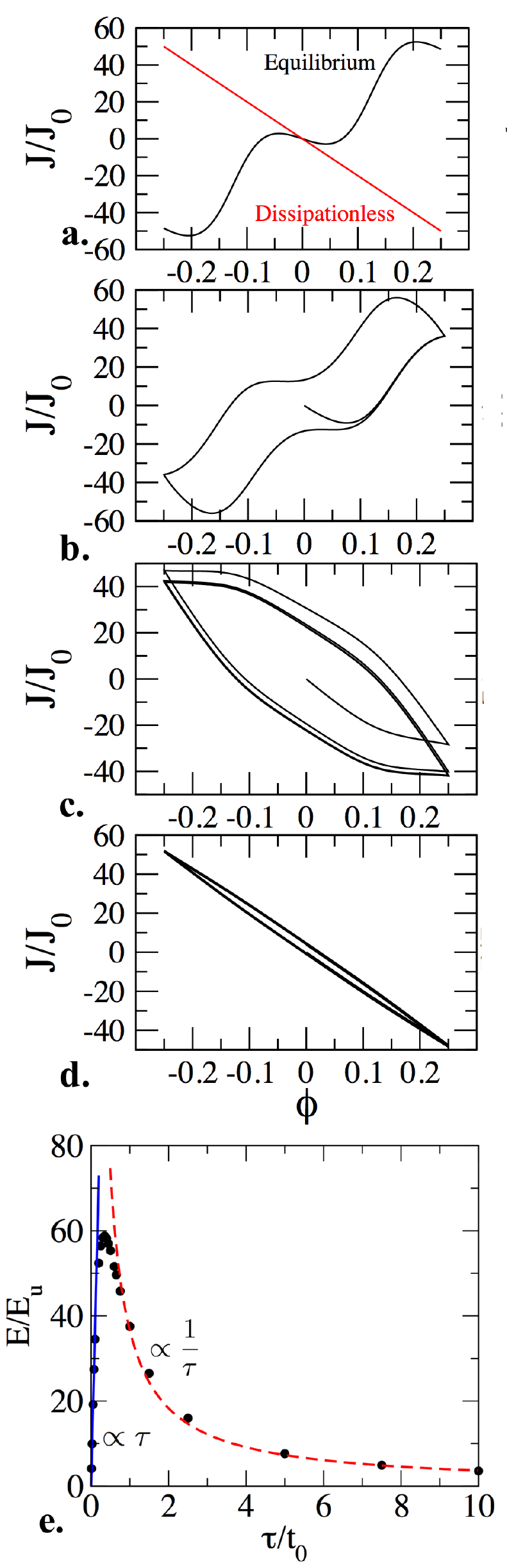}
\caption{Hysteresis loops and dissipated energy of spin-orbit coupled Fermi gases in a ring. (a-d) The total current, $J_{Tot}$,  as a function of the normalized flux, $\phi$, at $k_B T/E_u = 3$ with the spin-orbit coupling $\alpha/E_u = 0.75$. (a) The equilibrium (black) and dissipationless (red) currents for $N_\uparrow = N_\downarrow = 100$ fermions. The currents exhibit hysteresis loops as the flux is periodically modulated with relaxation times (b) $\tau/t_0 = 0.1$, (c) $\tau/t_0 = 1$, (d) and $\tau/t_0 = 10$. (e) Energy dissipated in one hysteresis loop as a function of the relaxation time $\tau$. The linear (blue) and inversely linear (red) dependence on $\tau$ in the strongly and weakly dissipative regimes similar to Fig.~\ref{Fig:Hyst_Area} are observable.}
\label{Fig:Hyst_SOC}
\end{figure}

The Hamiltonian of atoms traversing a 1D ring on the $xy$-plane acquires an extra coupling term from an additional artificial gauge field, which mimics an effective electric field and can be realized using laser to induce coupling of the spin to the momentum \cite{SOC_YJLin}. The effective coupling is dependent on the ratio of the momentum and energy of the laser. An artificial vector potential in the azimuthal direction $A\hat{\theta}$ is periodically modulating to drive an atomic current. The effective Hamiltonian in the presence of spin-orbit coupling is
\begin{equation}
H = \frac{1}{2m_f}(\vec{p}+\vec{A})^2 +\alpha [\hat{\sigma}\times (\vec{p}+\vec{A}) ]_z + V(r).
\end{equation}
The spin, represented by the vector $\hat{\sigma}$ consisting of Pauli matrices, couples to the momentum with coupling constant $\alpha$. Here a harmonic confining potential $V(r) \propto x^2$  is explicitly included, where $ x = (r-a)$ is the width of the confining potential with $a=L/(2\pi)$ being the ring radius \cite{SOC_HermHam}. Once the Hamiltonian has been formulated, the confining potential can be taken in the 1D limit $r \rightarrow a$ and $V(r) \rightarrow 0$. Using cylindrical coordinates where $\hat{\sigma}_r = \hat{\sigma}_x\cos(\theta)+\hat{\theta}_y\sin(\phi)$ and $\hat{\sigma}_\theta=\hat{\sigma}_y\cos(\theta) - \hat{\sigma}_x\sin(\theta)$, and using the previously defined relationship between $A$ and $\phi$, the Hamiltonian becomes
\begin{eqnarray}
H &=& -\frac{\hbar^2}{2m_f} \Big[ \frac{\partial^2}{\partial r^2} +\frac{1}{r}\frac{\partial}{\partial r}- \frac{1}{r^2}(i\frac{\partial}{\partial \theta} +\frac{\phi}{\phi_0})^2 \Big] \nonumber \\
& &  -\frac{\alpha}{r}\hat{\sigma_r}(i\frac{\partial}{\partial \theta} +\frac{\phi}{\phi_0})  +i\alpha\hat{\sigma}_{\theta}\frac{\partial}{\partial r} +V(r).
\end{eqnarray}

Separating the radial and azimuthal variables $H = H_r + H_{\theta}$ allows for the lowest radial mode, $R_0(r) \propto e^{-(r-a)^2/2}$ to be determined from the solution to the simple harmonic oscillator. After taking the limit $r\rightarrow a$, the Hamiltonian  $H_{1D} = \matrixel{R_0}{H_{\phi}}{R_0}$ is found to be
\begin{equation}
H_{1D} = E_u( i\frac{\partial}{\partial \theta} +\frac{\phi}{\phi_0} + \frac{\alpha}{2E_u}\hat{\sigma}_r)^2
\end{equation}
with $E_u = (2\pi\hbar)^2/(2m_f L^2)$. By solving $H\psi_n = E_n\psi_n$ with the following spinor
\begin{equation}
\psi_n^{\pm} = e^{in\theta} \begin{pmatrix}
a_n^{\pm}\\
e^{i\theta}b_n^{\pm}\\
\end{pmatrix},
\end{equation}
we obtain the energy eigenvalues
\begin{equation}
E_{n,\pm} =  E_u\Big((n-\phi) + \frac{1}{2}(1 \pm \sqrt{(\alpha/E_u)^2 +1})\Big)^2.
\end{equation}
Using the following relationship for the eigenfunction coefficients
$\frac{E_u}{\alpha}(1 \pm \sqrt{(\alpha/E_u)^2 +1})a_n^{\pm} = b_n^{\pm}$,
the eigenstates can be found with the trigonometric relationship $1/\cos(\theta) =  \sqrt{(\alpha/E_u)^2 +1}$. We obtain 
\begin{equation}
\psi_n^+ = e^{in\theta} \begin{pmatrix}
\cos(\frac{\theta}{2})\\
e^{i\theta}\sin(\frac{\theta}{2})\\
\end{pmatrix},~
\psi_n^- = e^{in\theta} \begin{pmatrix}
\sin(\frac{\theta}{2})\\
-e^{i\theta}\cos(\frac{\theta}{2})\\
\end{pmatrix}.
\end{equation}
An important feature is that the eigenstates have no dependence on the artificial magnetic flux $\phi$, meaning only the energy eigenvalues vary as the flux is turned on. This feature allows the system to be treated with the relaxation approximation, and the atomic current driven by a periodically modulating flux can be determined.

The current from a selected eigenstate can be found by $j_{n,\pm} = -\partial E_{n,\pm}/\partial\Phi$. Thus, 
\begin{equation}
j_{n,\pm} = J_0\Big((n-\phi) + \frac{1}{2}(1 \pm \sqrt{(\alpha/E_u)^2 +1})\Big).
\end{equation}
Here again $J_0 =(2\pi \hbar)/(m_f L^2)$. 
The current in the presence of RSOC and dissipation can be evaluated using the relaxation approximation in the same fashion as non-interacting fermions. We again implement the Fermi-Dirac statistics for the $E_{n,\pm}$ energy states in equilibrium.
\begin{equation}
f_{n,\pm,eq} = {1 \over e^{\beta(E_{n,\pm} - \mu)} + 1}.
\end{equation}
The chemical potential $\mu$ can be determined with a root finding method when evaluating $N_{tot} = N_{\uparrow}+N_{\downarrow}$ from the distribution functions,
$N_{tot} = \sum_n (f_{n,+}+f_{n,-})$.
The total current can be found by summing the current multiplied by the distribution function over all levels, $J_{Tot} = \sum_{n,\pm} f_{n,\pm}j_{n,\pm}$. The fully equilibrium current is obtained by using $f_{n,\pm,eq}$ in the evaluation. To determine the current in the presence of dissipation, we evaluate the distribution function using the relaxation approximation for $f_{n,\pm}$.
\begin{equation}
\frac{\partial f_{n,\pm}}{\partial t} = {-(f_{n,\pm}-f_{n,\pm,eq}) \over \tau}.
\end{equation}
 In Fig.~\ref{Fig:Hyst_SOC} the total current is plotted against the periodic, driving flux $\phi$ with fixed $\alpha/E_u = 0.75$. The period of the driving is again taken as $2t_0$. The persistent current (labeled as the equilibrium current) in the presence of RSOC decays rapidly with temperature, so we choose $k_B T = 3E_u$ in Fig.~\ref{Fig:Hyst_SOC}. Hysteresis loops are observable for $N_\uparrow = N_\downarrow = 100$ with spin-orbit interactions as shown in Figure~\ref{Fig:Hyst_SOC}(a-d). Similar hysteresis loops are observable for $N_\uparrow = N_\downarrow = 101$ with suitable parameter, too.

The dependence of energy dissipation on the relaxation time of the RSOC system also exhibits the universal behavior found in the noninteracting cases. Since the RSOC system can still be described in the framework of rate-dependent hysteresis, the linear and inverse-linear dependence of dissipated energy on the relaxation time holds true and is illustrated in  Figure~\ref{Fig:Hyst_SOC}(e). Therefore, the  RSOC induced by artificial gauge fields are also suitable for experimental measurements and observations of rate-dependent hysteresis in cold-atom systems.

\section{Conclusion}\label{sec:conclusion}
Hysteresis loops of atomic current have been demonstrated to exist for non-interacting fermions as well as fermions with spin-orbit coupling confined in a 1D ring potential with tunable dissipation, and similar rate-dependent hysteresis has be found in magnetic systems as well. The dissipated energy exhibits universal dependence on the relaxation time $\tau$. The universal behavior can be analyzed in the small and large $\tau$ regimes. The two-state paramagnet case is exactly solvable and the origin of the universal behavior arises from the response function possessing an identical structure as a damped harmonic oscillator in the massless limit. The observed dependence on the relaxation time is reminiscent to the dependence of the transition rate in Kramers transition rate theory of chemical reactions and the thermal conductance in 1D lattice on the friction coefficient. The observations illustrate striking similarities of dynamical phenomena across different research fields.

Hysteresis loops of noninteracting or spin-orbit coupled ultracold Fermi gases are promising because of recent advance in rapid loading of atoms \cite{Regal_AtomProd}, portable cold atom systems \cite{PortableAtoms}, and artificial gauge fields. Although cooling of fermionic systems is challenging, it is also a research field under active investigations. Usage of rate-independent hysteresis and related dynamic phenomena will bring opportunities for  atomtronic applications \cite{ChienNatPhys,LatticeMem}.

\textit{Acknowledgment} --- We thank Yu-Ju Lin, Kuei Sun, Michael Zwolak, and Roland Winston for stimulating discussions.

\bibliographystyle{apsrev4-1}
%

\end{document}